\documentclass{article}

\usepackage{microtype}
\usepackage{graphicx}
\usepackage{subfigure}
\usepackage{booktabs} 
\usepackage[colorlinks=true, citecolor=blue, urlcolor=blue, linkcolor=blue]{hyperref}

\usepackage[accepted]{icml2025}

\usepackage{amsmath}
\usepackage{amssymb}
\usepackage{mathtools}
\usepackage{amsthm}
\usepackage{paralist}

\usepackage[capitalize,noabbrev]{cleveref}

\theoremstyle{plain}

\theoremstyle{definition}

\theoremstyle{remark}

\usepackage[textsize=tiny]{todonotes}
\usepackage{enumitem}

\usepackage{pifont}

\icmltitlerunning{PiKV:  KV Cache Management System for MoE Architecture}

\begin{document}

\twocolumn[
\icmltitle{PiKV:  KV Cache Management System for MoE Architecture}
\icmlsetsymbol{equal}{*}



\begin{icmlauthorlist}
\icmlauthor{Dong Liu$^{*}$}{ucla,yale}
\icmlauthor{Yanxuan Yu}{columbia}
\icmlauthor{Ben Lengerich}{wisc}
\icmlauthor{Ying Nian Wu}{ucla}
\end{icmlauthorlist}

\icmlaffiliation{ucla}{University of California, Los Angeles}
\icmlaffiliation{yale}{Yale University}
\icmlaffiliation{columbia}{Columbia University}
\icmlaffiliation{wisc}{University of Wisconsin-Madison}

\icmlcorrespondingauthor{Dong Liu}{dong.liu@aya.yale.edu}

\icmlkeywords{Machine Learning, ICML}

\vskip 0.3in
]

\printAffiliationsAndNotice{}  

\begin{abstract}
As large-scale language models continue to scale up in both size and context length, the memory and communication cost of key-value (KV) cache storage has become a major bottleneck in multi-GPU and multi-node inference. While MoE-based architectures sparsify computation across experts, the corresponding KV caches remain dense and globally synchronized, resulting in significant overhead. 

We introduce \textbf{PiKV}, a parallel and distributed KV cache serving framework tailored for MoE architecture. PiKV leverages \textit{expert-sharded KV storage} to partition caches across GPUs, \textit{PiKV routing} to reduce token-to-KV access, and a \textit{PiKV Scheduling} to adaptively retain query-relevant entries. To further reduce memory usage, PiKV integrates \textit{PiKV Compression} modules the caching pipeline for acceleration.

PiKV is recently publicly available as an open-source software library: \href{https://github.com/NoakLiu/PiKV}{https://github.com/NoakLiu/PiKV}. PiKV is still a living project, aiming to become a comprehesive KV Cache management system for MoE Architectures.

\end{abstract}

\section{Introduction}

Large Language Models (LLMs) have become the foundation of modern AI applications, powering virtual assistants, code generation, document analysis, and multi-turn reasoning. With increasing demand for longer sequences and sparse expert models~\cite{bai2025qwen2,rajbhandari2020zero,openaiannouncement}, there is huge demand to deploy sparsely-gated Mixture-of-Experts (MoE) structures~\cite{lepikhin2020gshard,du2022glam} to reduce computation costs at scale.

However, serving such models introduces significant system-level challenges. During inference, each token generation requires attending to the entire KV cache from prior tokens. For a 7B-scale MoE model with 128K context and 16 experts, the full KV cache can occupy $>$24GB of memory and incur excessive communication latency across GPUs and nodes. Even with FlashAttention-style optimizations~\cite{dao2022flashattention}, the need to load and attend to dense KV structures becomes the dominant bottleneck, especially in autoregressive decoding.

Prior works~\cite{zhang2023h2o,gao2024cost} have shown that a small fraction of tokens contribute disproportionately to the final attention output, motivating selective cache access. Yet most methods either use static heuristics or ignore the underlying system cost of accessing KV entries across distributed compute nodes. In this work, we ask a deeper question: \emph{can we design a KV caching system that is both sparsity-aware and system-optimized for distributed MoE inference?}

\begin{figure}
    \centering
    \includegraphics[width=0.8\linewidth]{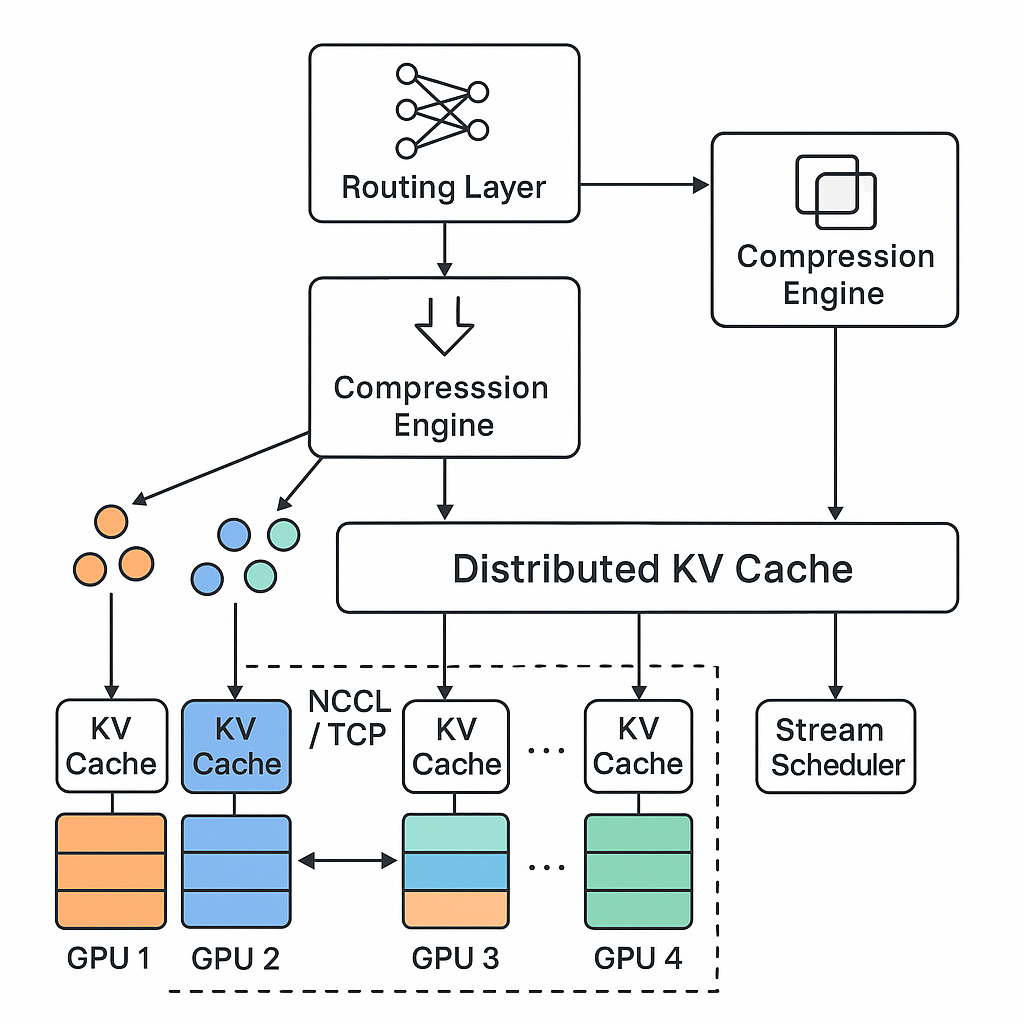}
    \caption{PiKV Framework}
    \label{fig:framework}
\end{figure}

We propose \textbf{PiKV}, a parallel distributed KV caching system tailored for sparse mixture of expert models training and inference. As shown in Figure~\ref{fig:framework}, PiKV includes three synergistic components: 
(1) an \textit{expert-sharded distributed KV cache} layout across multi-GPU or multi-node compute,
(2) a \textit{sparse expert routing layer} that dynamically selects top-$k$ experts per query,
and (3) an \textit{adaptive stream scheduler} that uses activity-based eviction to retain only high-utility KV entries. 

To further reduce memory and bandwidth cost, PiKV compresses KV representations using modular schemes such as LoRA~\cite{hu2022lora}, PyramidKV~\cite{cai2024pyramidkv}, and Duo~\cite{chen2024image}. We track metadata and usage patterns of each KV shard to guide eviction and cache streaming policies, enabling efficient inference under both static and streaming contexts.


\vspace{-10pt}
\begin{itemize}
    \setlength{\itemsep}{-3pt}
    \item We present a novel system architecture that combines sparse expert routing and distributed KV cache layout with query-aware streaming scheduling.
    \item We propose compression-aware KV caching, integrating multiple compression schemes and eviction policies into a unified system-level framework.
    \item We validate efficiency of PiKV in KV Cache Management of MoE Architectures, achieving significant improvements in memory, latency, and end-to-end generation efficiency.
\end{itemize}

\section{Related Work}

\subsection{Long-context LLMs}

With the rapid growth in model size and sequence length, long-context LLMs have gained significant attention. Architectures such as Transformer-XL~\cite{dai2019transformer}, Longformer~\cite{beltagy2020longformer}, and BigBird~\cite{zaheer2020bigbird} explored architectural sparsity to improve scalability. In practice, most LLMs adopt Rotary Position Embedding (RoPE)~\cite{su2024roformer}, which has been extended to longer contexts through rescaling~\cite{chen2023extending}. Notably, Yarn-LLaMA~\cite{peng2023yarn} expanded the 4K-token LLaMA-2 model to 32K and 128K tokens, respectively. Context length scaling techniques have pushed limits beyond 1M tokens~\cite{an2024training}. For long-document semantic understanding, hierarchical memory structures~\cite{liu2025hsgm} organize content into local segment graphs and a compact global summary, reducing worst-case semantic parsing complexity from $\mathcal{O}(N^2)$ to near-linear. Serving these long-context models introduces substantial KV cache pressure and decoding latency. PiKV addresses this by combining distributed KV cache design with token-level adaptive routing and scheduling.

\subsection{Sparse Expert Routing and KV Lookup}

MoE-based models~\cite{lepikhin2020gshard, fedus2022switch, du2022glam} reduce compute cost by activating a small subset of experts per token. However, inference-time memory and KV access patterns remain dense in most implementations. DeepSeek-V2~\cite{deepseekv2} and M6-MoE~\cite{zheng2023pit} explored dynamic expert selection and sparsity-aware routing. PiOne~\cite{lu2024not} and RingAttention~\cite{liu2023ring} explored multi-GPU KV alignment but did not optimize for token-expert selectivity. At the hardware level, CXL-SpecKV~\cite{cxlspeckv2026} leverages CXL interconnects and FPGA accelerators to disaggregate KV caches to remote memory, combining speculative prefetching with on-chip compression to reduce GPU memory pressure in datacenter deployments. Our work differs by tightly coupling token-level sparse routing with expert-sharded KV cache layouts and a lightweight backend gating network, reducing lookup and communication cost simultaneously.

\subsection{KV Compression and Stream-aware Scheduling}

To alleviate memory and bandwidth bottlenecks, several methods have been proposed to compress and truncate KV caches. H2O~\cite{zhang2023h2o} prioritizes tokens with high cumulative attention weights. FastGen~\cite{sheng2023flexgen} selects KV entries based on token type sensitivity. StreamingLLM~\cite{xiao2023streamingllm} introduce streaming sinks and proxy scoring methods to handle infinite-length texts. Quest~\cite{tang2024quest} and TOVA~\cite{he-etal-2025-hmt} further propose query-aware KV selection by evaluating relevance to the current query.

TinyServe~\cite{liu2025tinyserve} introduces query-aware page selection via bounding-box metadata to estimate per-block attention relevance, enabling selective KV loading with a fused CUDA kernel that integrates page scoring, sparse memory access, and masked attention in a single pass. LLMEasyQuant~\cite{liu2024llmeasyquant} further reduces KV footprint through scalable quantization tailored for parallel and distributed inference. Cache acceleration has also been studied in diffusion transformer models: FastCache~\cite{liu2025fastcache} employs learnable linear approximation to skip redundant cache computations, while AdaCorrection~\cite{liu2026adacorrection} applies adaptive offset correction to maintain generation fidelity under aggressive caching—insights that transfer to autoregressive LLM caching pipelines.

While effective, most methods either discard unused tokens too early or require full cache for scoring. In contrast, PiKV retains a unified KV cache layout with on-demand query-aware page loading, hierarchical compression (e.g., LoRA~\cite{hu2022lora}, PyramidKV~\cite{cai2024pyramidkv}), and streaming-aware eviction based on token activity scores. This provides a practical tradeoff between throughput, memory, and attention fidelity.

\subsection{Efficient ML Systems and Adaptive Planning}

Beyond KV cache management, broader system-level efforts have demonstrated that adaptive compute allocation improves efficiency across diverse ML workloads. Recent surveys comprehensively review efficiency techniques for large foundation models~\cite{liu2024designing}. GraphSnapShot~\cite{liu2025graphsnapshot} accelerates graph machine learning by decoupling the sampling and feature aggregation stages, showing that principled system co-design yields significant throughput gains. EchoRL~\cite{liu2025echorl} exploits experience replay to guide adaptive planning in reinforcement learning, illustrating how learned scheduling policies reduce wasted computation. Mt2St~\cite{liu2025mt2st} studies adaptive multi-task to single-task distillation, demonstrating that task-aware resource allocation can simplify complex inference pipelines—an insight directly applicable to expert-routing decisions in MoE serving. CoPrimeEEG~\cite{yu2026coprimeeeg} applies co-prime sub-Nyquist sampling with dual-branch reconstruction to EEG signals, illustrating how structured sparsity and adaptive reconstruction principles extend across data modalities. These works collectively motivate PiKV's philosophy of treating cache management as a dynamic, query-driven decision problem rather than a static policy.

\section{Methodology}
\label{sec:method}

The PiKV system is designed to rethink Key–Value (KV) cache management as a query-driven, memory–latency optimized process, tailored for sparse MoE inference at scale. In contrast to conventional cache systems that statically retain all past tokens, PiKV makes two fundamental shifts:

\begin{compactitem}
    \item \textbf{C1 (Sparsity-aware serving)}: Only a small set of experts and KV pages are relevant per query;
    \item \textbf{C2 (Resource-constrained scheduling)}: The memory and bandwidth budget must be dynamically partitioned across queries, experts, and streams.
\end{compactitem}
To this end, we decompose PiKV into four co-designed modules:
\textit{(i) distributed expert-sharded KV storage}, 
\textit{(ii) adaptive routing (\textit{PiKVRouting})}, 
\textit{(iii) modular compression (\textit{PiKVCompression})}, and 
\textit{(iv) query-aware stream scheduling (\textit{PiKVScheduling})},
with an optional \textit{(v) hardware-aware FPGA mapping} (\S\ref{sec:fpga}) that offloads metadata-heavy stages to reconfigurable accelerators.

All components are executed in an asynchronous pipeline orchestrated by a general decoding loop, as shown in Algorithm~\ref{alg:pikv-framework}. Each submodule operates independently but passes metadata to adjacent stages to inform decisions.

\begin{algorithm}[h]
\caption{General PiKV Execution Framework}
\label{alg:pikv-framework}
\begin{algorithmic}[1]
\STATE \textbf{Input:} query stream $\{q_t\}_{t=1}^T$, expert set $\mathcal{E}$, shard size $S$
\STATE Initialize: distributed cache $\mathcal{C}$, routing policy $\mathcal{R}$, scheduler $\mathcal{S}$, compressor $\mathcal{C}_{\text{cmp}}$
\FOR{$t=1$ to $T$}
    \STATE $g_t \gets \mathcal{R}(q_t)$ \hfill // PiKV Routing
    \STATE $K_t, V_t \gets f_{\text{enc}}(q_t)$ \hfill
    \FOR{expert $e \in g_t$}
        \STATE $s \gets \text{Shard}(t, e)$
        \STATE $(\hat{K}, \hat{V}) \gets \mathcal{C}_{\text{cmp}}(K_t, V_t)$ \hfill  // PiKV Compression
        \STATE $\mathcal{C}[e][s] \gets \text{Insert}((\hat{K}, \hat{V}), \text{metadata})$
    \ENDFOR
    \STATE $\mathcal{C} \gets \mathcal{S}(\mathcal{C}, q_t)$ \hfill // PiKV Scheduling
    \STATE $y_t \gets f_{\text{attn}}(q_t, \mathcal{C}[g_t])$
\ENDFOR
\end{algorithmic}
\end{algorithm}

We now describe each module and its underlying theoretical and system-level formulation.

\subsection{PiKV Expert-Sharded Storage}
\label{sec:kv}

Given a KV tensor pair $(K_t, V_t) \in \mathbb{R}^{d \times 2}$ at time $t$, the goal is to store these vectors in a distributed cache that minimizes redundancy and maximizes parallel retrieval. Unlike traditional schemes that replicate the full KV across $G$ GPUs, we assign tokens to shards via a hash function $h(t,e)$ and assign each shard to one GPU:

\[
s(t, e) = \left(t \bmod N_{\text{tok}}\right) \oplus \left(e \bmod N_{\text{exp}}\right).
\]

Each GPU stores only $\mathcal{O}(L/G + L/E)$ tokens, reducing per-device memory cost from $\mathcal{O}(EL)$.

\paragraph{Storage invariants.}
Each shard $s$ maintains a circular buffer of capacity $S$, so that insertions cost $\mathcal{O}(1)$ time and reallocation is avoided. If $(K_t, V_t)$ is compressed to $(\hat{K}_t, \hat{V}_t)$ of dimension $d'$, the per-shard memory is:

\[
\mathcal{M}_{s} = 2d'S = \frac{2dS}{\rho}, \quad \text{with } \rho = d / d'.
\]

Total memory per GPU is then:

\[
\mathcal{M}_{\text{kv}} = \frac{2d}{\rho}\left( \frac{L}{GS} + KS \right),
\]

where $K$ is the number of retained pages in PiKV scheduling.

\subsection{PiKV Routing}
\label{sec:routing}

PiKV Routing decides which experts $g_t \subseteq \mathcal{E}$ to activate for each query $q_t$. Formally, we define a routing function $\mathcal{R} : \mathbb{R}^d \to \{0,1\}^E$ satisfying $\|g_t\|_0 = k$.  PiKV supports multiple routing methods as in the following table \ref{tab:pikv-router-table}.



\begin{table}[h]
\centering
\caption{PiKV routing methods and their computational profiles ($E$ = experts).}
\label{tab:pikv-router-table}
\tiny
\begin{tabular}{llcc}
\toprule
\textbf{ID} & \textbf{Mechanism} & \textbf{Penalty Term} & \textbf{Cost} \\
\midrule
$\mathcal{R}_{\textbf{B}}$   & Base hash / round-robin      & —                                   & $\mathcal{O}(1)$             \\
$\mathcal{R}_{\textbf{T}}$   & TopK softmax                 & —                                   & $\mathcal{O}(E\!\log k)$     \\
$\mathcal{R}_{\textbf{LB}}$  & TopK + load balance          & $-\alpha(\mu_e-\bar{\mu})$          & $\mathcal{O}(E)$             \\
$\mathcal{R}_{\textbf{P}}$   & Cache-aware (PiKVRouter)     & $-\lambda\log\!\bigl(1+\text{miss}_e\bigr)$ & $\mathcal{O}(E)$     \\
$\mathcal{R}_{\textbf{E}}$   & Entropy-penalised LB (EPLB)  & $-\beta\,H(p_e)$                    & $\mathcal{O}(E)$             \\
$\mathcal{R}_{\textbf{A}}$   & RL-adaptive gating           & learned                             & $\mathcal{O}(k^2)$           \\
$\mathcal{R}_{\textbf{H}}$   & Hierarchical coarse$\!\to$fine & two-stage TopK                      & $\mathcal{O}(E+k\log k)$     \\
\bottomrule
\end{tabular}
\end{table}




\paragraph{Attention Complexity Reduction of PiKV Routing.}
Let $B$ be batch size, $L$ sequence length, $h$ head width, $E$ experts,
$k\!\ll\!E$ routed per token.

\[
\boxed{C_{\text{dense}} = B\,L\,h\,E},\qquad
\boxed{C_{\text{sparse}} = B\,L\,h\,k},
\quad\Longrightarrow\quad
\text{speed\!-\!up} = \frac{E}{k}.
\]

Memory traffic (Key–Value fetch, bytes):

\[
M_{\text{dense}} = 2B\,L\,h\,E,
\qquad
M_{\text{sparse}} = 2B\,L\,h\,k,
\quad
\text{relief} = \frac{E}{k}.
\]

Reuse-distance:

\[
\operatorname{RD}_{\text{dense}} \!=\! \frac{L}{E},
\qquad
\operatorname{RD}_{\text{sparse}} \!=\! \frac{L}{k}
\;\;\Longrightarrow\;\;
\text{hit-rate} \approx \frac{k}{E}.
\]


\subsection{PiKV Compression}
\label{sec:compression}

PiKV compression controls the space–fidelity trade-off for KV storage. Given $(K, V) \in \mathbb{R}^d \times \mathbb{R}^d$, a compressor $\mathcal{C}$ maps:

\[
\mathcal{C}(K, V) = (\hat{K}, \hat{V}) \in \mathbb{R}^{d'} \times \mathbb{R}^{d'}, \quad d' < d.
\]

We define the reconstruction error as:

\[
\epsilon = \frac{\|K - \mathcal{D}(\hat{K})\|_2}{\|K\|_2}, \quad \text{with decoder } \mathcal{D}.
\]

PiKV supports multiple compression methods as in the following table \ref{tab:pikv-compression-table}.

\begin{table}[h]
\centering
\caption{Analytic reconstruction bounds and asymptotic compression cost ($d$ = width, $r\!\ll\!d$ retained rank).}
\label{tab:pikv-compression-table}
\tiny
\begin{tabular}{llcc}
\toprule
\textbf{ID} & \textbf{Mechanism} & $\epsilon^{2}$ (squared error bound) & \textbf{Cost (Big-O)} \\
\midrule
$C_{\textbf{Lo}}$  & LoRA (rank $r$)           & $\;\displaystyle\sum_{i=r+1}^{d}\sigma_i^{2}$                & $\mathcal{O}(dr)$           \\[0.4ex]
$C_{\textbf{Lo}^+}$& LoRA$^{++}$               & $\;\bigl\|K-W_dW_uK-b\bigr\|_2^{2}$                          & $\mathcal{O}(dr)$           \\[0.4ex]
$C_{\textbf{Py}}$  & PyramidKV ($L$ levels)    & $\;\displaystyle\sum_{\ell=0}^{L-1}\frac{\|P^{(\ell)}K-K\|_2^{2}}{4^{\ell}}$ & $\mathcal{O}(d)$  \\[1.0ex]
$C_{\textbf{Ch}}$  & ChunkKV (block PCA)       & $\;\displaystyle\sum_{\text{blk}}\sum_{i>r}\sigma_i^{2}$     & $\mathcal{O}(dr)$           \\[0.8ex]
$C_{\textbf{SVD}}$ & Truncated SVD ($r$)       & $\;\displaystyle\sum_{i>r}\sigma_i^{2}$                      & $\mathcal{O}(d^{2}r)$\footnotemark \\[0.4ex]
$C_{\textbf{F}}$   & FastV (crop to $r$)       & $\;\|K_{r:d}\|_2^{2}$                                         & $\mathcal{O}(d)$            \\[0.4ex]
$C_{\textbf{Dis}}$ & Distillation (offline)    & $\;\mathrm{KL}(q_{\text{teach}}\!\parallel\!q_{\text{stud}})$ & $\mathcal{O}(d\,r)$\\[0.4ex]
$C_{\textbf{Pr}}$  & Structured Pruning        & $\;\displaystyle\sum_{j\in\mathcal{Z}}K_j^{2}$               & $\mathcal{O}(d)$            \\[0.4ex]
\bottomrule
\end{tabular}
\footnotetext{Full SVD is offline; at inference only the $O(dr)$ projection is executed.}
\end{table}










\paragraph{Compression-Aware Latency of PiKV Compression.}
\emph{Variables:}
$d$ full width, $d' = d/\rho$ compressed width ($\rho>1$),  
$k$ experts/query, $B$ tokens/batch,  
$\beta$ HBM bandwidth (B/s), $\gamma$ core throughput (B/s),  
$\eta\le 2$ decode factor.

\vspace{-0.4em}
\begin{align}
T_{\text{read}}   &= \frac{2d'kB}{\beta}
                  = \frac{2dkB}{\rho\beta},                             \tag{1} \\[0.2em]
T_{\text{decode}} &= \frac{\eta d'kB}{\gamma}
                  = \frac{\eta dkB}{\rho\gamma},                        \tag{2} \\[0.2em]
T_{\text{step}}   &= T_{\text{read}} + T_{\text{decode}}
                  = \frac{dkB}{\rho}\!\Bigl(\tfrac{2}{\beta}+\tfrac{\eta}{\gamma}\Bigr).\tag{3}
\end{align}

\paragraph{Speed-up.}
For two compression ratios $\rho_1<\rho_2$,
\[
\mathrm{Speedup}(\rho_1\!\rightarrow\!\rho_2)
  =\frac{T_{\text{step}}(\rho_1)}{T_{\text{step}}(\rho_2)}
  =\frac{\rho_2}{\rho_1}. \tag{4}
\]




Higher $\rho$ linearly reduces both read and decode time until
$T_{\text{decode}}\!\approx\!T_{\text{read}}$, after which the gain plateaus.

\subsection{PiKV Scheduling}
\label{sec:scheduling}

PiKV Scheduler implements dynamic retention of cached KV pages under bounded memory. Instead of static eviction rules, PiKV formulates scheduling as a per-page scoring problem, where each entry $i$ is assigned a scalar utility score $u_i$ based on features such as attention intensity, recency of access, and reuse patterns. PiKV supports multiple scheduling methods as in following table \ref{tab:pikv-scheduling-formal}


\begin{table}[h]
\centering
\caption{
Summary of PiKV scheduling strategies.
Notation: $a_i$ = attention, $r_i$ = recency, $f_i$ = frequency, $t_i$ = age, $\phi_j(i)$ = feature scores, $\theta$ = eviction threshold, $\eta$ = hit-rate.
$\checkmark$ = adaptive threshold, $\times$ = fixed.
}
\label{tab:pikv-scheduling-formal}
\small
\begin{tabular}{llc}
\toprule
\textbf{ID} & \textbf{Scheduling Methods $u_i$} & \textbf{Adaptive} \\
\midrule
$\mathcal{S}_{\text{H2O}}$     & $u_i = a_i$                                                 & $\times$ \\
$\mathcal{S}_{\text{SL}}$      & $u_i = \mathbb{I}[t_i > \tau]$                              & $\times$ \\
$\mathcal{S}_{\text{QUEST}}$   & $u_i = \mathrm{MLP}_\theta\big([K_i, V_i]\big)$             & $\checkmark$ \\
$\mathcal{S}_{\text{Flex}}$    & $u_i = \mathcal{M}_{\text{plan}}(t_i)$                      & $\times$ \\
$\mathcal{S}_{\text{LRU}}$     & $u_i = -r_i$                                                & $\times$ \\
$\mathcal{S}_{\text{LRU+}}$    & $u_i = -r_i + \lambda \cdot f_i$                            & $\times$ \\
$\mathcal{S}_{\text{AdaKV}}$   & $u_i = \sum_j \alpha_j \phi_j(i)$,\quad $\theta \leftarrow \theta + \gamma(\eta^* - \eta)$ & $\checkmark$ \\
$\mathcal{S}_{\text{Duo}}$     & $u_i = \sum_{\ell=1}^{L} a_i^{(\ell)}$                      & $\checkmark$ \\
\bottomrule
\end{tabular}
\end{table}




\paragraph{Memory Usage of PiKV.}
We analyze the total per-GPU memory consumption $\mathcal{M}_{\text{total}}$ of PiKV under compressed KV storage and bounded scheduling. Let:
\begin{itemize}
    \item $d$: original hidden size of each KV vector;
    \item $\rho = d / d'$: compression ratio, where $d'$ is the reduced dimensionality;
    \item $L$: number of cached tokens per expert globally;
    \item $G$: number of GPUs (i.e., KV shards);
    \item $S$: circular buffer size (in tokens) per expert shard;
    \item $K$: number of active cache pages selected by the scheduler per GPU.
\end{itemize}

The total memory per GPU decomposes into two parts:
\begin{align*}
\mathcal{M}_{\text{token}} &= \frac{2d'}{G} \cdot \frac{L}{S}, &&\text{(sharded token buffer)} \\
\mathcal{M}_{\text{page}}  &= 2d' \cdot K \cdot S,            &&\text{(scheduled page buffer)}
\end{align*}

Summing the two and replacing $d' = d/\rho$ yields:
\[
\mathcal{M}_{\text{total}} = \mathcal{M}_{\text{token}} + \mathcal{M}_{\text{page}} = \frac{2d}{\rho} \left( \frac{L}{GS} + KS \right).
\]

To minimize $\mathcal{M}_{\text{total}}$ with respect to $S$, we take the derivative:
\begin{align*}
\frac{\partial \mathcal{M}_{\text{total}}}{\partial S} &= -\frac{2dL}{\rho G S^2} + \frac{2dK}{\rho}, \\
\text{set } \frac{\partial \mathcal{M}_{\text{total}}}{\partial S} = 0 \quad &\Rightarrow \quad
S^* = \sqrt{ \frac{L}{KG} }.
\end{align*}

Therefore, the optimal buffer size $S^*$ trades off between sharding granularity and reuse coverage. Substituting back:
\[
\mathcal{M}_{\text{total}}^\ast = \frac{4d}{\rho} \sqrt{ \frac{KL}{G} }.
\]

This closed-form provides a practical design rule for setting shard capacity $S$ to minimize GPU memory cost under fixed compression $\rho$, token budget $L$, and scheduler retention $K$.

\subsection{Hardware-Aware FPGA Mapping}
\label{sec:fpga}

While \S\ref{sec:kv}--\S\ref{sec:scheduling} optimize KV management in GPU-centric clusters, datacenter MoE serving increasingly faces a \emph{memory wall}: HBM capacity scales more slowly than context length and expert count. PiKV's modular pipeline is deliberately structured so that metadata-intensive stages can be lifted to a \textbf{hardware-aware} FPGA tier without changing the logical API seen by the GPU attention kernels.

\paragraph{System topology.}
Table~\ref{tab:fpga-general} shows the \textbf{PiKV-FPGA} stack. The host GPU runs $f_{\text{enc}}$ and $f_{\text{attn}}$; an FPGA SmartNIC (e.g., AMD Alveo U55C / Intel Agilex) sits on a CXL Type-3 link~\cite{cxlspeckv2026} to expanded DDR and exposes a 32\,B MMIO command queue to the driver. KV \emph{payloads} live in disaggregated memory; the FPGA keeps only metadata, scores, and codec weights on chip.

\begin{figure}[t]
\centering
\fbox{\parbox{0.92\linewidth}{\centering\vspace{1.8em}
\textbf{GPU} $\xrightarrow{\text{MMIO}}$ \textbf{PiKV-CTRL}
$\rightarrow$ \{$\mathcal{D}$, ScoreFuse, $\mathrm{Codec}_\rho$, DMA\}
$\xrightarrow{\text{CXL.mem}}$ \textbf{DDR pool}\\
\vspace{0.4em}
GPU $\xleftarrow{\text{PCIe/CXL}}$ packed $\{(\hat{K},\hat{V}, \mathrm{idx})\}_{i \in \mathcal{P}_t}$
\vspace{1.8em}}}
\caption{PiKV-FPGA topology: control on FPGA, KV bodies in CXL-attached memory.}
\label{fig:fpga-topo}
\end{figure}

\paragraph{Module-to-engine mapping.}
Table~\ref{tab:fpga-general} maps each PiKV method variant (\S\ref{sec:routing}--\S\ref{sec:scheduling}) to a reconfigurable engine; Table~\ref{tab:fpga-map} summarizes the shared on-chip modules and state.

\begin{table}[t]
\centering
\caption{Hardware-aware mapping of PiKV methods to FPGA engines ($T_{\mathrm{fpga}}$: on-chip cycles).}
\label{tab:fpga-general}
\tiny
\begin{tabular}{@{}llcc@{}}
\toprule
\textbf{ID} & \textbf{Method} & \textbf{$\mathcal{E}_{\cdot}$} & \textbf{$T_{\mathrm{fpga}}$} \\
\midrule
\multicolumn{4}{@{}l}{\textit{Routing}} \\
$\mathcal{R}_{\textbf{B}}$ & hash / round-robin & $\mathcal{D}$ lookup & $\mathcal{O}(1)$ \\
$\mathcal{R}_{\textbf{T}}$ & TopK softmax & ScoreFuse + radix Top-$k$ & $\mathcal{O}(E\!\log k)$ \\
$\mathcal{R}_{\textbf{LB}}$ & load balance & ScoreFuse $-\alpha(\mu_e\!-\!\bar{\mu})$ & $\mathcal{O}(E)$ \\
$\mathcal{R}_{\textbf{P}}$ & cache-aware & ScoreFuse $-\lambda\log(1\!+\!m_e)$ + prefetch & $\mathcal{O}(E)$ \\
$\mathcal{R}_{\textbf{E}}$ & entropy-penalised LB & ScoreFuse $-\beta H(p_e)$ & $\mathcal{O}(E)$ \\
$\mathcal{R}_{\textbf{H}}$ & hierarchical TopK & 2-stage ScoreFuse & $\mathcal{O}(E\!+\!k\!\log k)$ \\
\midrule
\multicolumn{4}{@{}l}{\textit{Compression}} \\
$C_{\textbf{Lo}}$ & LoRA ($r$) & rank-$r$ URAM matvec & $\mathcal{O}(dr)$ \\
$C_{\textbf{Py}}$ & PyramidKV & multi-level $\mathrm{Codec}_\rho$ & $\mathcal{O}(d)$ \\
$C_{\textbf{Ch}}$ & ChunkKV & block-PCA engine & $\mathcal{O}(dr)$ \\
$C_{\textbf{F}}$ & FastV & tail crop + pad & $\mathcal{O}(d)$ \\
$C_{\textbf{Pr}}$ & structured prune & sparse mask gen. & $\mathcal{O}(d)$ \\
\midrule
\multicolumn{4}{@{}l}{\textit{Scheduling}} \\
$\mathcal{S}_{\text{H2O}}$ & $u_i{=}a_i$ & max-reduce attention & $\mathcal{O}(K)$ \\
$\mathcal{S}_{\text{SL}}$ & sliding window & age comparator & $\mathcal{O}(1)$ \\
$\mathcal{S}_{\text{QUEST}}$ & MLP scorer & DSP MLP + $\theta$ BRAM & $\mathcal{O}(dK)$ \\
$\mathcal{S}_{\text{LRU}}$ & $u_i{=}{-}r_i$ & recency sort network & $\mathcal{O}(K\!\log K)$ \\
$\mathcal{S}_{\text{AdaKV}}$ & multi-feature + $\theta$ & fused $\sum_j\alpha_j\phi_j$; MMIO $\theta$ & $\mathcal{O}(K)$ \\
$\mathcal{S}_{\text{Duo}}$ & layer-sum attention & $L$-way acc.\ on $a_i^{(\ell)}$ & $\mathcal{O}(LK)$ \\
\bottomrule
\end{tabular}
\end{table}

\begin{table}[t]
\centering
\caption{Shared PiKV-FPGA modules ($\Gamma$: page table, $m_e$: miss count).}
\label{tab:fpga-map}
\small
\begin{tabular}{@{}ccc@{}}
\toprule
\textbf{Module} & \textbf{$\mathcal{E}_{\cdot}$} & \textbf{SRAM} \\
\midrule
$\mathcal{C}$      & $\mathcal{D}$ + gather     & $\Gamma\!:\!(t,e)\!\mapsto\!a$ \\
$\mathcal{R}$      & $\mathrm{Top}\text{-}k\!\circ\!\mathcal{L}$ & $\{\mu_e, m_e\}$ \\
$\mathcal{C}_{\mathrm{cmp}}$ & $\mathrm{Codec}_\rho$    & $\{W, \sigma\}$ \\
$\mathcal{S}$      & $u_i \gtrless \theta$      & $\{(r_i, f_i)\}$ \\
\bottomrule
\end{tabular}
\end{table}

\paragraph{Resource budget.}
On-chip memory is allocated as follows (bit-widths in parentheses):
\begin{align*}
\mathrm{BRAM}_{\Gamma} &= E\cdot S \cdot (32{+}48), \\
\mathrm{BRAM}_{\mathrm{meta}} &= k K S \cdot (16{+}16{+}16), \\
\mathrm{URAM}_{W} &= d \cdot r \;\text{(LoRA)}.
\end{align*}
For a representative MoE serving tile ($E{=}64$, $S{=}256$, $k{=}4$, $K{=}16$, $d{=}128$), $\mathrm{BRAM}_{\Gamma}{\approx}176$\,KB and $\mathrm{BRAM}_{\mathrm{meta}}{\approx}48$\,KB, fitting a single U55C SLR with headroom for ScoreFuse and DMA buffers.

\paragraph{Latency and bandwidth.}
End-to-end FPGA latency per token is
\[
T_{\text{fpga}} = T_{\mathrm{route}} + k\bigl(T_{\Gamma} + K(T_{\mathrm{ddr}} + T_{\mathrm{codec}})\bigr),
\]
with $T_{\mathrm{route}}{=}\lceil E/16\rceil/f_{\text{fpga}}$, $T_{\Gamma}{=}2/f_{\text{fpga}}$, and $T_{\mathrm{ddr}}{=}2d'/B_{\mathrm{mem}}$. Host–GPU traffic is
\[
B_{\text{step}} \approx \frac{2 k d' |\mathcal{P}_t|}{\rho_{\text{link}}} + k\log E,
\]
where $\rho_{\text{link}}$ is on-chip compression and $|\mathcal{P}_t|{\le}kK$. When $B_{\text{step}} \ll 2dL$, the GPU sees only $\mathcal{C}[g_t]$ and stays off the metadata critical path. Adaptive $\mathcal{S}_{\text{AdaKV}}$ updates $\theta$ in BRAM every $\Delta$ tokens via MMIO without re-synthesis.

\section{Conclusion}




We present \textbf{PiKV}, a parallel and distributed KV cache management framework optimized for sparsely activated MoE-based large language models. PiKV introduces a KV cache management system for MoE, including sparse expert routing, cache compression, and stream-aware scheduling. This architecture rethinks KV caching not only as passive memory storage, but as a dynamic, query-driven retrieval system.


\href{https://github.com/NoakLiu/PiKV}{PiKV} is a living project for scalable MoE serving, aiming to bridge MoE sparsity and efficient system design optimization. Future work will explore online adaptation, hierarchical memory tiers, and integration with training-time sparsity strategies for end-to-end efficient large model deployment with MoE architecture.

\bibliography{main}

@article{bai2025qwen2,
  title={Qwen2. 5-vl technical report},
  author={Bai, Shuai and Chen, Keqin and Liu, Xuejing and Wang, Jialin and Ge, Wenbin and Song, Sibo and Dang, Kai and Wang, Peng and Wang, Shijie and Tang, Jun and others},
  journal={arXiv preprint arXiv:2502.13923},
  year={2025}
}

@inproceedings{rajbhandari2020zero,
  title={Zero: Memory optimizations toward training trillion parameter models},
  author={Rajbhandari, Samyam and Rasley, Jeff and Ruwase, Olatunji and He, Yuxiong},
  booktitle={SC20: International Conference for High Performance Computing, Networking, Storage and Analysis},
  pages={1--16},
  year={2020},
  organization={IEEE}
}

@article{peng2023yarn,
  title={Yarn: Efficient context window extension of large language models},
  author={Peng, Bowen and Quesnelle, Jeffrey and Fan, Honglu and Shippole, Enrico},
  journal={arXiv preprint arXiv:2309.00071},
  year={2023}
}

@misc{openaiannouncement,
  title={Gpt-4 technical report},
  author={Achiam, Josh and Adler, Steven and Agarwal, Sandhini and Ahmad, Lama and Akkaya, Ilge and Aleman, Florencia Leoni and Almeida, Diogo and Altenschmidt, Janko and Altman, Sam and Anadkat, Shyamal and others},
  journal={arXiv preprint arXiv:2303.08774},
  year={2023}
}

@article{lepikhin2020gshard,
  title={Gshard: Scaling giant models with conditional computation and automatic sharding},
  author={Lepikhin, Dmitry and Lee, HyoukJoong and Xu, Yuanzhong and Chen, Dehao and Firat, Orhan and Huang, Yanping and Krikun, Maxim and Shazeer, Noam and Chen, Zhifeng},
  journal={arXiv preprint arXiv:2006.16668},
  year={2020}
}

@inproceedings{du2022glam,
  title={Glam: Efficient scaling of language models with mixture-of-experts},
  author={Du, Nan and Huang, Yanping and Dai, Andrew M and Tong, Simon and Lepikhin, Dmitry and Xu, Yuanzhong and Krikun, Maxim and Zhou, Yanqi and Yu, Adams Wei and Firat, Orhan and others},
  booktitle={International conference on machine learning},
  pages={5547--5569},
  year={2022},
  organization={PMLR}
}

@article{dao2022flashattention,
  title={Flashattention: Fast and memory-efficient exact attention with io-awareness},
  author={Dao, Tri and Fu, Dan and Ermon, Stefano and Rudra, Atri and R{\'e}, Christopher},
  journal={Advances in neural information processing systems},
  volume={35},
  pages={16344--16359},
  year={2022}
}

@article{zhang2023h2o,
  title={H2o: Heavy-hitter oracle for efficient generative inference of large language models},
  author={Zhang, Zhenyu and Sheng, Ying and Zhou, Tianyi and Chen, Tianlong and Zheng, Lianmin and Cai, Ruisi and Song, Zhao and Tian, Yuandong and R{\'e}, Christopher and Barrett, Clark and others},
  journal={Advances in Neural Information Processing Systems},
  volume={36},
  pages={34661--34710},
  year={2023}
}

@inproceedings{gao2024cost,
  title={$\{$Cost-Efficient$\}$ large language model serving for multi-turn conversations with $\{$CachedAttention$\}$},
  author={Gao, Bin and He, Zhuomin and Sharma, Puru and Kang, Qingxuan and Jevdjic, Djordje and Deng, Junbo and Yang, Xingkun and Yu, Zhou and Zuo, Pengfei},
  booktitle={2024 USENIX Annual Technical Conference (USENIX ATC 24)},
  pages={111--126},
  year={2024}
}

@inproceedings{sheng2023flexgen,
  title={Flexgen: High-throughput generative inference of large language models with a single gpu},
  author={Sheng, Ying and Zheng, Lianmin and Yuan, Binhang and Li, Zhuohan and Ryabinin, Max and Chen, Beidi and Liang, Percy and R{\'e}, Christopher and Stoica, Ion and Zhang, Ce},
  booktitle={International Conference on Machine Learning},
  pages={31094--31116},
  year={2023},
  organization={PMLR}
}

@article{hu2022lora,
  title={Lora: Low-rank adaptation of large language models.},
  author={Hu, Edward J and Shen, Yelong and Wallis, Phillip and Allen-Zhu, Zeyuan and Li, Yuanzhi and Wang, Shean and Wang, Lu and Chen, Weizhu and others},
  journal={ICLR},
  volume={1},
  number={2},
  pages={3},
  year={2022}
}

@article{cai2024pyramidkv,
  title={Pyramidkv: Dynamic kv cache compression based on pyramidal information funneling},
  author={Cai, Zefan and Zhang, Yichi and Gao, Bofei and Liu, Yuliang and Liu, Tianyu and Lu, Keming and Xiong, Wayne and Dong, Yue and Chang, Baobao and Hu, Junjie and others},
  journal={arXiv preprint arXiv:2406.02069},
  year={2024}
}

@inproceedings{chen2024image,
  title={An image is worth 1/2 tokens after layer 2: Plug-and-play inference acceleration for large vision-language models},
  author={Chen, Liang and Zhao, Haozhe and Liu, Tianyu and Bai, Shuai and Lin, Junyang and Zhou, Chang and Chang, Baobao},
  booktitle={European Conference on Computer Vision},
  pages={19--35},
  year={2024},
  organization={Springer}
}

@article{dai2019transformer,
  title={Transformer-xl: Attentive language models beyond a fixed-length context},
  author={Dai, Zihang and Yang, Zhilin and Yang, Yiming and Carbonell, Jaime and Le, Quoc V and Salakhutdinov, Ruslan},
  journal={arXiv preprint arXiv:1901.02860},
  year={2019}
}

@article{beltagy2020longformer,
  title={Longformer: The long-document transformer},
  author={Beltagy, Iz and Peters, Matthew E and Cohan, Arman},
  journal={arXiv preprint arXiv:2004.05150},
  year={2020}
}

@article{zaheer2020bigbird,
  title={Big bird: Transformers for longer sequences},
  author={Zaheer, Manzil and Guruganesh, Guru and Dubey, Kumar Avinava and Ainslie, Joshua and Alberti, Chris and Ontanon, Santiago and Pham, Philip and Ravula, Anirudh and Wang, Qifan and Yang, Li and others},
  journal={Advances in neural information processing systems},
  volume={33},
  pages={17283--17297},
  year={2020}
}

@article{su2024roformer,
  title={Roformer: Enhanced transformer with rotary position embedding},
  author={Su, Jianlin and Ahmed, Murtadha and Lu, Yu and Pan, Shengfeng and Bo, Wen and Liu, Yunfeng},
  journal={Neurocomputing},
  volume={568},
  pages={127063},
  year={2024},
  publisher={Elsevier}
}

@article{chen2023extending,
  title={Extending context window of large language models via positional interpolation, 2023},
  author={Chen, Shouyuan and Wong, Sherman and Chen, Liangjian and Tian, Yuandong},
  journal={URL https://arxiv. org/abs/2306.15595}
}

@article{an2024training,
  title={Training-free long-context scaling of large language models},
  author={An, Chenxin and Huang, Fei and Zhang, Jun and Gong, Shansan and Qiu, Xipeng and Zhou, Chang and Kong, Lingpeng},
  journal={arXiv preprint arXiv:2402.17463},
  year={2024}
}

@article{fedus2022switch,
  title={Switch transformers: Scaling to trillion parameter models with simple and efficient sparsity},
  author={Fedus, William and Zoph, Barret and Shazeer, Noam},
  journal={Journal of Machine Learning Research},
  volume={23},
  number={120},
  pages={1--39},
  year={2022}
}

@article{deepseekv2,
  title={Deepseek-v2: A strong, economical, and efficient mixture-of-experts language model},
  author={Liu, Aixin and Feng, Bei and Wang, Bin and Wang, Bingxuan and Liu, Bo and Zhao, Chenggang and Dengr, Chengqi and Ruan, Chong and Dai, Damai and Guo, Daya and others},
  journal={arXiv preprint arXiv:2405.04434},
  year={2024}
}

@inproceedings{zheng2023pit,
  title={Pit: Optimization of dynamic sparse deep learning models via permutation invariant transformation},
  author={Zheng, Ningxin and Jiang, Huiqiang and Zhang, Quanlu and Han, Zhenhua and Ma, Lingxiao and Yang, Yuqing and Yang, Fan and Zhang, Chengruidong and Qiu, Lili and Yang, Mao and others},
  booktitle={Proceedings of the 29th Symposium on Operating Systems Principles},
  pages={331--347},
  year={2023}
}

@article{lu2024not,
  title={Not all experts are equal: Efficient expert pruning and skipping for mixture-of-experts large language models},
  author={Lu, Xudong and Liu, Qi and Xu, Yuhui and Zhou, Aojun and Huang, Siyuan and Zhang, Bo and Yan, Junchi and Li, Hongsheng},
  journal={arXiv preprint arXiv:2402.14800},
  year={2024}
}

@article{liu2023ring,
  title={Ring attention with blockwise transformers for near-infinite context},
  author={Liu, Hao and Zaharia, Matei and Abbeel, Pieter},
  journal={arXiv preprint arXiv:2310.01889},
  year={2023}
}

@article{tang2024quest,
  title={Quest: Query-aware sparsity for efficient long-context llm inference},
  author={Tang, Jiaming and Zhao, Yilong and Zhu, Kan and Xiao, Guangxuan and Kasikci, Baris and Han, Song},
  journal={arXiv preprint arXiv:2406.10774},
  year={2024}
}

@inproceedings{he-etal-2025-hmt,
    title = "{HMT}: Hierarchical Memory Transformer for Efficient Long Context Language Processing",
    author = "He, Zifan  and
      Cao, Yingqi  and
      Qin, Zongyue  and
      Prakriya, Neha  and
      Sun, Yizhou  and
      Cong, Jason",
    editor = "Chiruzzo, Luis  and
      Ritter, Alan  and
      Wang, Lu",
    booktitle = "Proceedings of the 2025 Conference of the Nations of the Americas Chapter of the Association for Computational Linguistics: Human Language Technologies (Volume 1: Long Papers)",
    month = apr,
    year = "2025",
    address = "Albuquerque, New Mexico",
    publisher = "Association for Computational Linguistics",
    url = "https://aclanthology.org/2025.naacl-long.410/",
    pages = "8068--8089",
    ISBN = "979-8-89176-189-6"
}

@article{xiao2023streamingllm,
      title={Efficient Streaming Language Models with Attention Sinks}, 
      author={Guangxuan Xiao and Yuandong Tian and Beidi Chen and Song Han and Mike Lewis},
      year={2024},
      eprint={2309.17453},
      archivePrefix={arXiv},
      primaryClass={cs.CL},
      url={https://arxiv.org/abs/2309.17453}, 
}

@inproceedings{liu2025tinyserve,
author = {Liu, Dong and Yu, Yanxuan},
title = {TinyServe: Query-Aware Cache Selection for Efficient LLM Serving},
year = {2025},
isbn = {9798400720352},
publisher = {Association for Computing Machinery},
address = {New York, NY, USA},
url = {https://doi.org/10.1145/3746027.3758181},
doi = {10.1145/3746027.3758181},
booktitle = {Proceedings of the 33rd ACM International Conference on Multimedia},
pages = {12529–12537},
numpages = {9},
location = {Dublin, Ireland},
series = {MM '25}
}

@inproceedings{liu2025hsgm,
    title = "{HSGM}: Hierarchical Segment-Graph Memory for Scalable Long-Text Semantics",
    author = "Liu, Dong  and
      Yu, Yanxuan",
    editor = "Frermann, Lea  and
      Stevenson, Mark",
    booktitle = "Proceedings of the 14th Joint Conference on Lexical and Computational Semantics (*SEM 2025)",
    month = nov,
    year = "2025",
    address = "Suzhou, China",
    publisher = "Association for Computational Linguistics",
    url = "https://aclanthology.org/2025.starsem-1.26/",
    doi = "10.18653/v1/2025.starsem-1.26",
    pages = "328--337",
    ISBN = "979-8-89176-340-1"
}

@inproceedings{cxlspeckv2026,
  author = {Liu, Dong and Yu, Yanxuan},
  title = {CXL-SpecKV: A Disaggregated FPGA Speculative KV-Cache for Datacenter LLM Serving},
  year = {2026},
  isbn = {9798400720796},
  publisher = {Association for Computing Machinery},
  address = {New York, NY, USA},
  url = {https://doi.org/10.1145/3748173.3779188},
  doi = {10.1145/3748173.3779188},
  booktitle = {Proceedings of the 2026 ACM/SIGDA International Symposium on Field Programmable Gate Arrays},
  pages = {56–66},
  numpages = {11},
  location = {USA},
  series = {FPGA '26}
}

@article{liu2024designing,
  title={Designing large foundation models for efficient training and inference: A survey},
  author={Liu, Dong and Yu, Yanxuan and Wang, Yite and Wu, Jing and Wan, Zhongwei and Alinejad, Sina and Lengerich, Benjamin and Wu, Ying Nian},
  journal={arXiv preprint arXiv:2409.01990},
  year={2024}
}

@article{liu2024llmeasyquant,
      title={LLMEasyQuant: Scalable Quantization for Parallel and Distributed LLM Inference}, 
      author={Dong Liu and Yanxuan Yu},
      year={2025},
      eprint={2406.19657},
      archivePrefix={arXiv},
      primaryClass={cs.LG},
      url={https://arxiv.org/abs/2406.19657}, 
}

@inproceedings{
liu2025graphsnapshot,
title={GraphSnapShot: A System for Graph Machine Learning Acceleration},
author={Dong Liu and Yanxuan Yu},
booktitle={Machine Learning for Computer Architecture and Systems 2025},
year={2025},
url={https://openreview.net/forum?id=KeHes2SVxs}
}

@inproceedings{
liu2025echorl,
title={Echo{RL}: Learning to Plan through Experience for Efficient Reinforcement Learning},
author={Dong Liu and Yanxuan Yu and Ying Nian Wu},
booktitle={The 5th Workshop on Mathematical Reasoning and AI at NeurIPS 2025},
year={2025},
url={https://openreview.net/forum?id=34d8tFWoOR}
}

@article{liu2025fastcache,
  title={Fastcache: Fast caching for diffusion transformer through learnable linear approximation},
  author={Liu, Dong and Yu, Yanxuan and Zhang, Jiayi and Li, Yifan and Lengerich, Ben and Wu, Ying Nian},
  journal={arXiv preprint arXiv:2505.20353},
  year={2025}
}

@article{liu2026adacorrection,
  title={AdaCorrection: Adaptive Offset Cache Correction for Accurate Diffusion Transformers},
  author={Liu, Dong and Yu, Yanxuan and Lengerich, Ben and Wu, Ying Nian},
  journal={arXiv preprint arXiv:2602.13357},
  year={2026}
}

@article{yu2026coprimeeeg,
  title={CoPrimeEEG: CRT-Guided Dual-Branch Reconstruction from Co-Prime Sub-Nyquist EEG},
  author={Yu, Yanxuan and Liu, Dong and Wu, Ying Nian},
  journal={bioRxiv},
  pages={2026--02},
  year={2026},
  publisher={Cold Spring Harbor Laboratory}
}

@inproceedings{liu2025mt2st,
  title={Mt2st: Adaptive multi-task to single-task learning},
  author={Liu, Dong and Yu, Yanxuan},
  booktitle={Proceedings of the 1st Workshop on Multimodal Augmented Generation via Multimodal Retrieval (MAGMaR 2025)},
  pages={79--89},
  year={2025}
}
\bibliographystyle{icml2025}

\end{document}